\documentclass[twocolumn,superscriptaddress,prl]{revtex4}

\usepackage{graphicx}
\usepackage{epsfig}
\usepackage{verbatim}
\usepackage{bm}

\newcommand{\gammabar}{\ensuremath\gamma\kern-0.53em-}

\usepackage{times}
\usepackage{amsmath}
\usepackage{amsfonts}
\usepackage{amssymb}
\usepackage{graphicx}
\usepackage{pst-all}
\usepackage{leftidx}

\usepackage{tikz}

\makeatletter
\def\l@subsubsection#1#2{}
\makeatother

\begin{document}

\title{Charge $2e/3$ superconductivity and topological degeneracies without localized zero modes in \\
bilayer fractional quantum Hall states}
\author{Maissam Barkeshli}
\affiliation{Station Q, Microsoft Research, Santa Barbara, California 93106-6105, USA}

\begin{abstract}
It has been recently shown that non-Abelian defects with localized parafermion zero modes can arise
in conventional Abelian fractional quantum Hall (FQH) states. Here we propose an alternate route
to creating, manipulating, and measuring topologically protected degeneracies in bilayer
FQH states coupled to superconductors, without the creation of localized parafermion zero modes. We focus mainly on
electron-hole bilayers, with a $\pm 1/3$ Laughlin FQH state in each layer, with boundaries that are
proximity-coupled to a superconductor. We show that the superconductor induces charge $2e/3$ quasiparticle-pair
condensation at each boundary of the FQH state, and that this leads to (1) topologically protected degeneracies
that can be measured through charge sensing experiments and (2) a fractional charge $2e/3$
AC Josephson effect. We demonstrate that an analog of non-Abelian braiding is possible,
despite the absence of a localized zero mode. We discuss several practical advantages of this
proposal over previous work, and also extensions to electron-electron bilayers at $\nu = 2/3 + 1/3$ coupled to
superconductivity, and to electron-hole bilayers with only interlayer tunneling.
\end{abstract}

\maketitle

A profound feature of topologically ordered states of matter is the existence of topologically protected degeneracies \cite{nayak2008}.
These are degeneracies in the ground state spectrum of a given Hamiltonian for which no local operator can distinguish the
topologically distinct states, thus rendering them robust to local perturbations. Topological degeneracies are predicted to
occur when a topological phase is defined on a space with non-trivial topology, such as a torus \cite{wen1989,wen1990b},
when the system hosts non-Abelian quasiparticle excitations \cite{Moore1991,wen1991prl,nayak1996,bonderson2011}, and
in topological superconductors when Majorana zero modes are bound to cores of vortices or the ends of one-dimensional wires \cite{read2000,kitaev2001,fu2008,sau2010,alicea2010,lutchyn2010,oreg2010,alicea2012review}.

It has recently been shown theoretically that a wide class of non-Abelian defects can be realized in conventional
Abelian fractional quantum Hall (FQH) states, such as the Laughlin FQH states,
by suitably controlling the tunneling processes along the edge modes. In particular, it was shown that in bilayer
FQH states, domain walls between regions of inter-layer and intra-layer tunneling
can give rise to a novel type of non-Abelian defect that generalizes Majorana zero modes by yielding
$m$ states per pair of defects, where the integer $m$ depends on the filling fraction $\nu$ of the FQH state \cite{barkeshli2012a}.
These non-Abelian defects were subsequently shown to also arise as lattice defects in certain lattice spin
models\cite{you2012}, extending \cite{bombin2010}. They were then also shown to arise in
FQH systems coupled to superconductors, at domain walls between regions of normal and Andreev backscattering
at the edge \cite{clarke2013, lindner2012, cheng2012}, where they were referred to as
parafermion \cite{fendley2012}, or fractionalized Majorana, zero modes.
The general theory of these defects \cite{kitaev2012,barkeshli2013genon,barkeshli2013defect2,teo2014,barkeshli2014SDG},
a number of experimental proposals in FQH systems \cite{barkeshli2014,barkeshli2014deg,clarke2014}, other exotic phenomena
at the boundary of FQH states and superconductivity \cite{barkeshli2015fqhsc}, and theoretical suggestions
for more exotic non-Abelian states of matter \cite{vaezi2013,mong2014,vaezi2013b,vaezi2014fib,barkeshli2015kitaev}, have since been developed.
\begin{figure}
	\centering
	\includegraphics[width=3.4in]{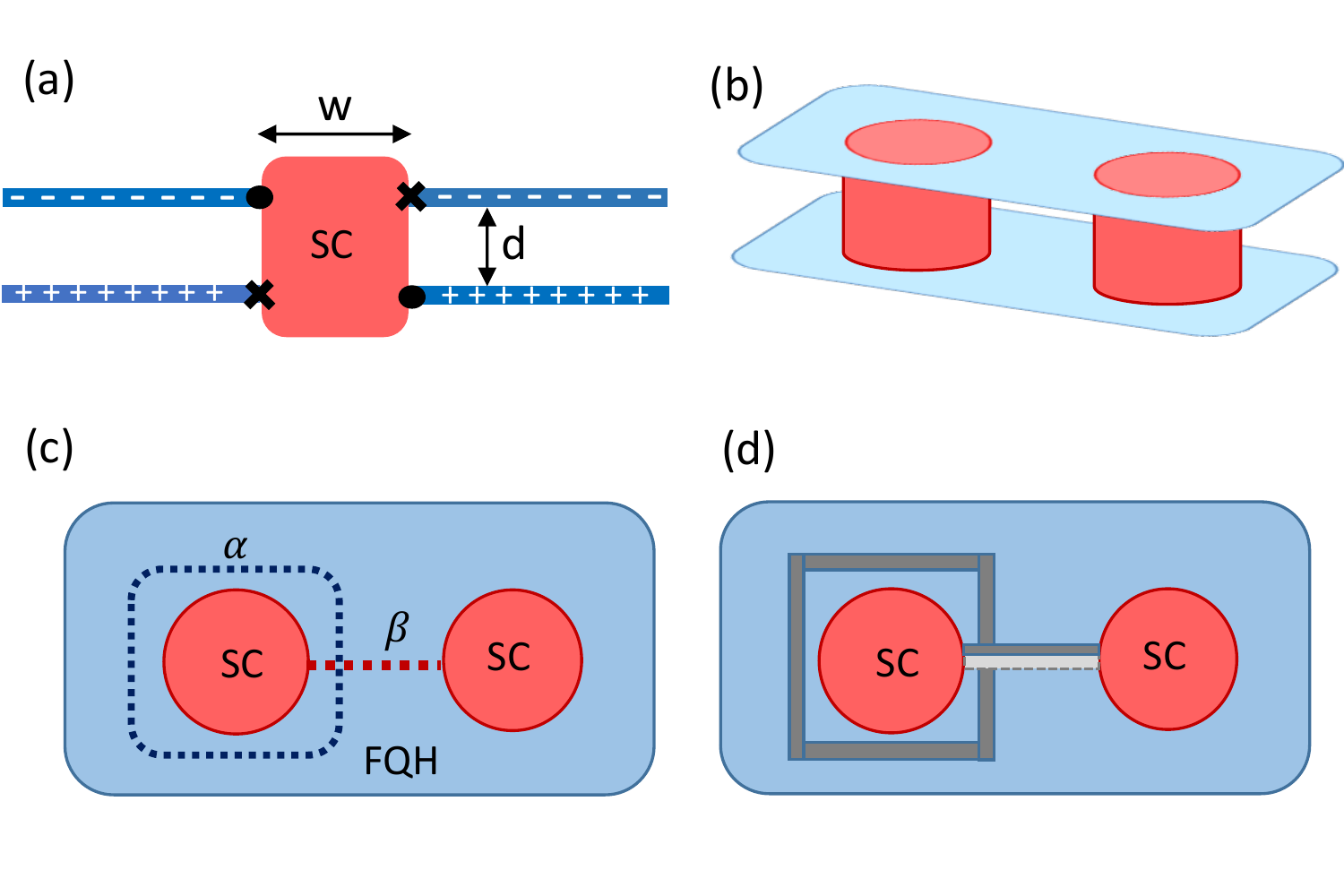}
	\caption{\label{fig1} (a) Cross-section. Bilayer electron-hole FQH system, with a $|\nu| = 1/3$ Laughlin FQH state in each layer.
Cooper pairs can tunnel between the counterpropagating FQH edge modes and the superconductor (red). This can happen in the regime
$d < \xi_{\text{sc}}$, where $\xi_{\text{sc}}$ is the superconducting coherence length and $l_B$ is the magnetic length.
(b) 3D view of two regions filled with superconductor.
(c) Top view. Readout of topological states can be performed by measuring the charge, in units of $2e/3$, in the region enclosed by the loop $\alpha$.
(d) Top and bottom gates (dark and light gray, respectively) to enhance quasiparticle tunneling along $\alpha$ or $\beta$. }
\end{figure}

In this paper, we propose a path to realizing topological degeneracies in bilayer FQH systems in contact with superconductivity
without the existence of a localized parafermion zero mode. We consider an electron-hole bilayer
FQH system, with a $1/3$ Laughlin FQH state in each layer. We show that the superconductor can induce charge $2e/3$ condensation
along the boundary, giving rise to $3$ topological ground states for each boundary component. Our proposal
allows the possibility of detecting the topological states entirely through charge-sensing measurements, requires only one type
of backscattering term along the FQH edge, and allows large interlayer separation $d$ and width $w$ between the edge states,
thus simplifying the situation as compared with the proposals in \cite{clarke2013, lindner2012,cheng2012,barkeshli2014, mong2014,barkeshli2014deg}.
Extending recent work \cite{barkeshli2016mcg}, we further demonstrate that despite the absence of a
point-like localized zero mode, the system does support an analog of non-Abelian braiding, which can be induced
through the use of carefully designed top and bottom electric gate configurations. These results are also of
theoretical interest for the problem of defining modular transformations in fermionic topological phases.

We note that the concept of boundary-induced topological degeneracies underlie some $Z_2$ surface code proposals \cite{fowler2012}
and has also been discussed abstractly in \cite{wang2015}. The main contribution here is to discuss a concrete physical platform
and to demonstrate the theoretical possibility of a topologically robust non-Abelian Berry phase.

\it Setup \rm -- Let us consider an electron-hole bilayer quantum Hall system, such that the electrons and holes are at
filling fraction $\nu = \pm 1/3$, i.e. $\nu_{\text{total}} = -1/3 + 1/3$. For $d/l_B \gg 1$, where $d$ and $l_B$
are the layer separation and magnetic length, respectively, the system can realize an independent $1/3$ Laughlin
state in each layer \footnote{We note that $n+1/3$ Laughlin states, for integer $n$, are also sufficient for our purpose.}
\footnote{$d/l_B < 1$ may also be sufficient, although this depends on microscopic
details of the interactions}. Since the two layers possess opposite charge carriers (holes vs. electrons),
the chiralities of the Laughlin states in each layer are reversed with respect to each other.

For concreteness, we further consider the case where the two layers carry opposite
spin, thus effectively realizing a fractional quantum spin Hall (FQSH) state.
In the case where both layers carry the same spin, the superconductivity-induced effects described below
will require a superconductor with either spin triplet pairing or with strong spin-orbit coupling.

The FQSH state described above can potentially be realized in a variety of experimental systems.
These include electron-hole bilayers in semiconductor quantum well heterostructures with opposite $g$-factor
in the two layers \cite{shabani}, or in double-layer graphene with a thin dielectric separating the two
layers and top/bottom gates to separately control the electron/hole densities in each layer. The spin structure
of the graphene Landau levels \cite{abanin2006} can then potentially support a FQSH state at $\nu = -1/3 + 1/3$ \cite{young}.
Twisted bilayer graphene is also a possible candidate, as it has been shown to realize an integer QSH state \cite{maher2013,sanchez2016}.
Recent experiments have demonstrated advances in coupling superconductors to semiconductor 2DEGS \cite{rokhinson2015,shabani2015}, e.g. through epitaxial
growth \cite{shabani2015} and to graphene in the quantum Hall regime \cite{finkelstein2015}.

The edge of the bilayer FQH system described above now consists of counterpropagating chiral Luttinger liquids,
described by the following Lagrangian density \cite{wen1995}:
\begin{align}
\mathcal{L}_0 = \sum_{I = 1,2} \frac{(-1)^I 3 }{4\pi} \partial_t \phi_I \partial_x \phi_I - V_{IJ} \partial_x \phi_I \partial_x \phi_J.
\end{align}
$\phi_I$, for $I = 1, 2$ are counterpropagating chiral scalar fields satisfying a periodicity condition $\phi_I \sim \phi_I + 2\pi$.
The electron annihilation operator on each edge is given by $\Psi_{1\uparrow} \sim e^{i 3 \phi_{1\uparrow}}$, $\Psi_{2\downarrow} \sim e^{i 3 \phi_{3\downarrow}}$,
the charge $e/3$ quasiparticle in each layer is given by $e^{i \phi_I}$, and the charge density in each layer is
$\rho_I = \frac{(-1)^I}{2\pi} \partial_x \phi_I$. $V_{IJ}$ is a positive-definite matrix encoding the velocities and
interactions between the edge modes. Upon quantization of the above theory,
the chiral fields satisfy the commutation relation $[\phi_I(x), \phi_J(y)] = \delta_{IJ} (-1)^I \frac{\pi}{3} \text{sgn}(x-y)$.
Finally, it is understood that the $\phi_I$ carry opposite spin, although it is not explicitly labelled.

Given such an edge, one could consider two types of electron backscattering terms between the two edge modes:
\begin{align}
\mathcal{L}_{t} = -M \Psi_{1\uparrow}^\dagger \Psi_{2 \downarrow} - \Delta \Psi_{1\uparrow} \Psi_{2\downarrow} + H.c.
\end{align}
In terms of the bosonized variables, $\mathcal{L}_t = - M \cos(3 (\phi_1 - \phi_2)) - |\Delta| \cos(3 (\phi_1 + \phi_2) + \Phi )$,
where $\Delta = |\Delta| e^{i \Phi}$. The first term corresponds to normal electron backscattering between the two edge modes;
since the edge modes carry opposite spin, such a term requires strong spin-orbit coupling, or magnetism, to allow
the required spin-flip ($M$ is taken to be real for simplicity). The second term can arise when the boundary of the system is coupled to a conventional
$s$-wave superconductor, and corresponds to a process where a Cooper pair from the superconductor hops onto the counterpropagating chiral edge modes.
It requires $d < \xi_{\text{sc}}$, where $\xi_{\text{sc}}$ is the superconducting coherence length.
Ref. \cite{clarke2013,lindner2012,cheng2012} considered domain walls between regions where the normal and
Andreev backscattering terms were dominant. Below we will consider the case where only the Andreev backscattering
term is dominant along the entire boundary (later we will comment on the case where only the normal backscattering term is
dominant).

\it Charge 2e/3 superconductivity \rm --
With the appropriate interactions $V_{IJ}$, the Andreev backscattering term above can be made relevant in the
renormalization-group sense. Alternatively, even if the tunneling term is irrelevant in the RG sense, we can
consider the bare strength of $|\Delta|$ to be on the order of the gap of the FQH state. In such a case,
where $\Delta$ is the dominant term, the argument of the cosine is pinned to a constant value corresponding to one of its three
inequivalent minima: $(\phi_1 + \phi_2) = 2\pi n/3 - \Phi/3$, for $n = 0,1,2$, thus inducing a finite energy gap along the boundary.
Formally, this is possible because $[(\phi_1(x)+\phi_2(x)), \phi_1(y) + \phi_2(y)] = 0$. In particular, this implies
\begin{align}
\langle n | e^{i (\phi_1 + \phi_2)}|m \rangle = \delta_{mn} e^{-i \Phi/3 + 2\pi n/3},
\end{align}
where $|n \rangle$ for $n = 0,1,2$ are the three states associated with the three minima of the cosine.
The operator $e^{\pm i(\phi_1 + \phi_2)}$ is the creation/annihilation operator for a charge
$2e/3$ quasiparticle on the edge of the system, consisting of an $e/3$ quasiparticle from each layer.
Thus the boundary of the FQH system forms a charge $2e/3$ superconductor, as condensation of charge
$2e/3$ quasiparticle pairs has been induced.

The condensation of charge $2e/3$ quasiparticles along the boundary of the system implies that, in the absence of
a mesoscopic charging energy, they can be added to the edge at no additional energy cost, similar to how Cooper pairs
can be added to a conventional superconductor at no energy cost. In contrast, adding charge $e/3$ quasiparticles costs a
finite energy, as this causes a kink in $\phi_1 + \phi_2$ due to the commutation relation
$[\phi_I(x), (\phi_1(x) + \phi_2(y))] = (-1)^I i \frac{\pi}{3} \text{sgn}(x-y)$.

While charge $2e/3$ quasiparticles are condensed, the flux quantization of the superconductor through a hole bounded by the
$2e/3$ condensate is still in units of $hc/2e$ \footnote{This follows
from the fact that flux $3hc/2e$ is clearly allowed from the charge $2e/3$ pairing, while $hc/e$ is always allowed by gauge invariance, implying that $hc/2e$ must
also be allowed.}.

\begin{figure}
	\centering
	\includegraphics[width=3.4in]{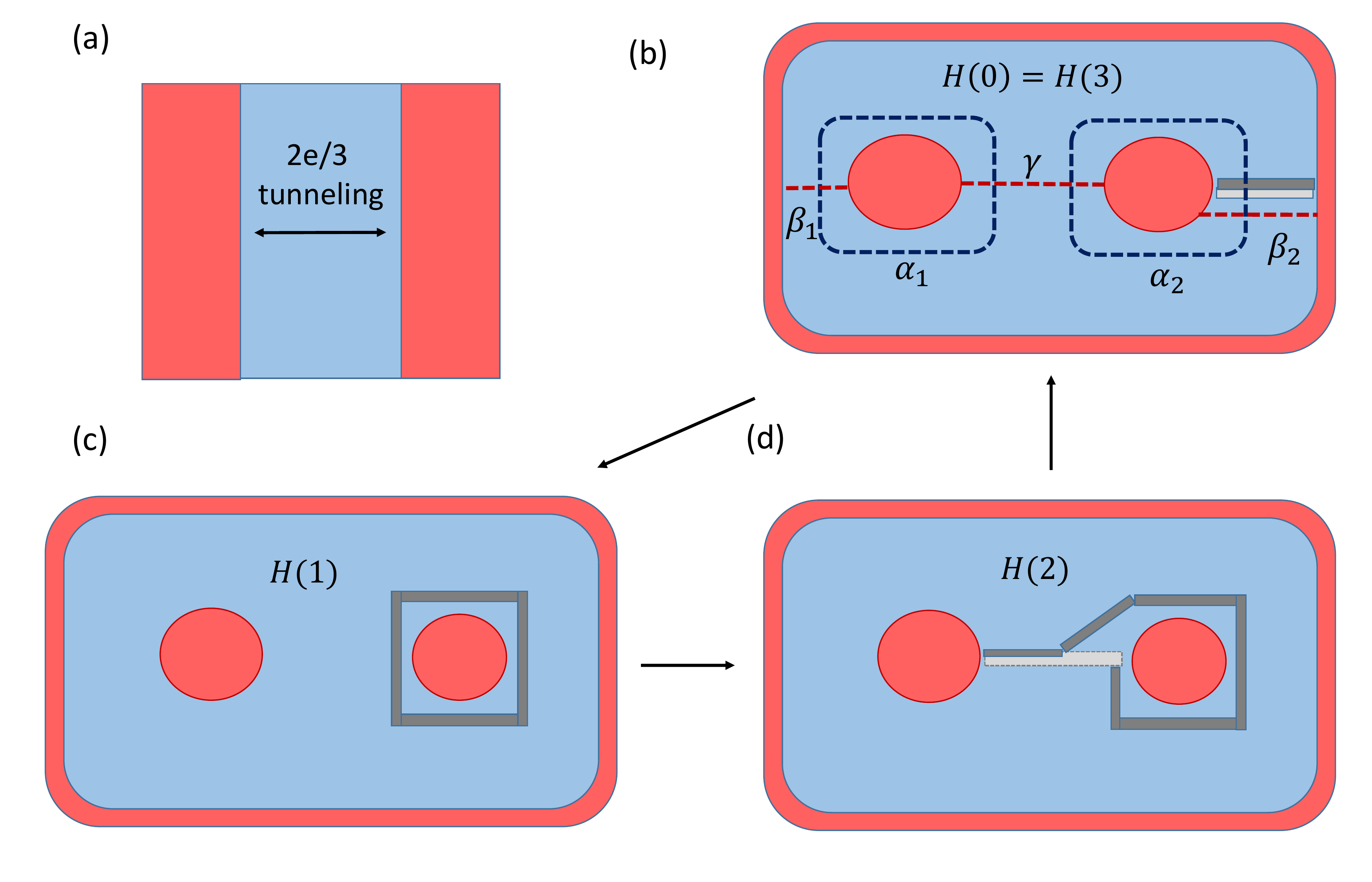}
	\caption{\label{fig2} (a) S - FQH -S geometry for fractional Josephson effect. (b)-(d) Evolution of $H(\tau)$ for $\tau = 0,1,2,3$
Systems includes three disconnected boundaries coupled to superconductors (red). Dashed lines indicate loops/lines of interest. Dark grey bars indicate active top gates,
light grey bars indicate active bottom gates.   }
\end{figure}

\it Charge $2e/3$ Josephson effect \rm -- Let us consider a S-FQH-S junction (see Fig. \ref{fig2}(a)).
Since charge $2e/3$ quasiparticles can be added to boundary at zero energy
cost, we can consider at arbitrarily low energies the process where a charge $2e/3$ quasiparticle
tunnels coherently from one boundary to another. In the presence of a voltage difference across the S-FQH-S jnction,
this induces the tunneling Hamiltonian
\begin{align}
H_{J} = \int dx t(x) e^{i 2e V t /3} e^{i (\phi_{A1} + \phi_{A2})} e^{-i (\phi_{B1} + \phi_{B2})} + H.c.,
\end{align}
where $\phi_{BI}$ and $\phi_{AI}$ denote the fields from the two boundaries, respectively.
Since $e^{i (\phi_{A1} + \phi_{A2})}$ and $e^{i (\phi_{B1} + \phi_{B2})}$ are condensed, we can replace them by their expectation
values, and arrive at the Hamiltonian $H_J = \int dx |t(x)| \cos( (\Phi_A - \Phi_B + 2 e V t)/3 - \theta(x))$, where $t(x) = |t(x)|e^{i\theta(x)}$.
This implies a bulk AC current $I \propto \sin( (\Phi_A - \Phi_B)/3 - \theta(x) + 2e V t/3)$ with a fractional Josephson frequency
$\omega_J = 2e V/3$.

This is in contrast to the case with parafermion zero modes studied in \cite{clarke2013,lindner2012,cheng2012}, which
yielded a charge $2e/6$ Josephson effect, due to the fact that single electrons could also tunnel
coherently between the localized zero modes. The fractional Josephson effect described here is due to tunneling along the entire edge,
not only from one point to another as in \cite{clarke2013,lindner2012,cheng2012}, and is more similar to the proposal of
\cite{senthil2001} to diagnose fractionalization in the cuprates through a charge $e$ Josephson effect.

\it Topological ground state degeneracy \rm --
To further understand the topological degeneracies, consider two disconnected boundaries (Fig. \ref{fig1}b-c),
each with characteristic size $L$ and separation between them also of order $L$.
The low-lying states of the system can be written as $|n_1,n_2\rangle$, for integer $n_1, n_2$, where
$2e n_1/3$ and $2e n_2/3$ is the amount of charge associated to each boundary region, respectively.
Fixing the overall charge of the system means that the state is fully specified by the difference $n_1 - n_2$.
Different states have an energy splitting $\propto 1/L$, arising from electrostatic charging energies.

Now let us consider the two disconnected boundaries to be coupled to the \it same \rm superconductor. We take the superconductor
to be of size $L_{\text{sc}}$, while the boundaries are still considered to have size and separation betweeen them of order $L$.
In this case, it is no longer meaningful to discuss the charge associated to each boundary region, as the charge is now completely delocalized
throughout the entire superconductor. However, there is still a meaning to the topological, or anyonic, charge associated
to each boundary region. Specifically, the number of charge $2e/3$ quasiparticles assigned to each boundary region, modulo three,
is still a well-defined quantity, due to the fractional statistics of the quasiparticles.

The system has two basic types of Wilson operators, $W_1(\alpha)$ and $W_2(\beta)$  (see Fig. \ref{fig1}(c)).
$W_1(\alpha)$ creates a charge $e/3$, $-e/3$ quasiparticle-quasihole pair in the top layer, tunnels one of
them around the loop $\alpha$ (in the top layer), and reannihilates them.
$W_2(\beta)$ tunnels a charge $2e/3$ quasiparticle, which consists of a an $e/3$ quasiparticle in each layer,
from one boundary condensate to the other. These two processes leave the system in the ground state subspace
in the limit $L, L_{\text{sc}} \rightarrow \infty$. All other operators that also leave the system in the ground state subspace
can be obtained through products of the above non-local string operators. Importantly, the fractional statistics of the quasiparticles implies
$W_1(\alpha) W_2(\beta) = W_2(\beta) W_1(\alpha) e^{2\pi i /3}$.
The above algebra guarantees that the system has three ground states, as the ground state subspace must form an irreducible representation
of the above algebra. The basis states can be labelled
$|n \rangle_{\beta}$, such that $W_1(\alpha) |n \rangle_{\beta} = | (n + 1) \% 3 \rangle_\beta$, and
$W_2(\beta) |n \rangle_\beta = e^{2\pi i n/3} |n\rangle_\beta$. An alternate basis $|n \rangle_\alpha$ can be defined:
$W_2(\beta) |n \rangle_{\alpha} = |(n + 1) \% 3 \rangle_\alpha$, and $W_1(\alpha) |n \rangle_\alpha = e^{2\pi i n/3} |n\rangle_\alpha$,
where $|n \rangle_\alpha = \sum_m e^{2\pi i m n/3} |n \rangle_\beta$.
No local operator can distinguish the states $|n \rangle$ -- only the non-local operators $W$ can -- which
is why the degeneracy is topologically protected.

For $n_b$ disconnected gapped boundaries, each characterized by a charge $2e/3$ condensate, one obtains
$n_b - 1$ distinct copies of the above algebra, leading to $3^{n_b - 1}$ ground states.

For a finite system, virtual tunneling corresponding to $W_1(\alpha)$ and $W_2(\beta)$ yields an
exponentially small splitting in the degeneracy, proportional to $e^{-L/\xi}$, as described below.

\it State preparation and detection \rm--
Let us consider now the case where the superconductors associated with the two boundary regions are disconnected (Fig. \ref{fig1}(b)-(c)).
The charge associated to each superconducting region is then well-defined and can in principle be measured experimentally,
for example by capacitively coupling to a single-electron transistor (SET), similar to other fractional charge sensing
experiments in FQH systems \cite{venkatachalam2011}. In other words, the topological
states can be read out in the basis $|n \rangle_\alpha$ by measuring the fractional charge, in units of $2e/3$,
in the region enclosed by $\alpha$ (see Fig. \ref{fig1}(c)).

The effective Hamiltonian in the low energy topological subspace takes the form:
\begin{align}
H_{\text{eff}} = \frac{Q_1^2}{2C} + \frac{Q_2^2}{2C} - t_1 W_1(\alpha) - t_{2} W_2(\beta) + H.c.
\end{align}
Here, $Q_1$ and $Q_2$ are the total charge in first and second superconducting boundary regions, respectively, and we have
included a capacitance $C$ to ground for each superconductor (for simplicity we ignore cross-capacitances).
The first two terms represent the charging energy for each superconducting region.

$t_1$ is the net amplitude for creating a quasiparticle-quasihole pair of charge $e/3$, $-e/3$
out of the ground state, tunneling the charge $e/3$ quasiparticle clockwise around any loop that is topologically equivalent to $\alpha$,
and reannihilating them, summed together with the net amplitude for the reverse process in the bottom layer (the effect of
$W_1(\alpha)$ is reversed if the $e/3$ quasiparticle is tunneling in the bottom layer). In general, $t_1 \propto e^{-c_1 L_\alpha/\xi}$,
where $L$ is length of the smallest loop that is topologically equivalent to $\alpha$, $\xi \sim l_B$ is the correlation length
of the system, which is set by the energy gap of the FQH state, and $c_1$ is a constant term that depends on the energy gap
to creating charge $e/3$ quasiparticles. $t_2$ is the net amplitude for creating a pair of charge $e/3$ quasiparticles out of the one of
the boundary condensates and annihilating it at the other boundary condensate, and thus $t_2 \propto e^{-c_2 L_\beta/\xi}$, where
$L_\beta$ is the minimal distance between the two boundaries, and $c_2$ is a constant that depends
on the gap to creating a pair of charge $2e/3$ quasiparticles, one in each layer, in the bulk of the FQH state. In principle, the effective
Hamiltonian also contains higher order processes, such as tunneling charge $m e/3$ quasiparticles around the loop $\alpha$, or
tunneling charge $2m e/3$ quasiparticles from one boundary to another, although these will be exponentially suppressed relative
to the terms kept above, as the energy gaps for creating such quasiparticles is expected to be larger.

Ignoring the charging energies, when $t_1 \gg t_2$, the ground state is an eigenstate of $W_1(\alpha)$, and thus corresponds to a definite $|n \rangle_\alpha$.
When $t_2 \gg t_1, e^2/C$, the ground state is instead an eigenstate of $W_2(\beta)$, which means that it
must be of the form $|a\rangle_{\beta} = \sum_a e^{2\pi i a n  /3} |n\rangle_\alpha$.

The magnitudes of $t_1$ and $t_2$ can be tuned experimentally through the use of electric gates. Applying a local electric potential
to the FQH fluid can reduce the energy gap $\epsilon_{e/3}$ to creating a charge $e/3$ quasiparticle, while simultaneously increasing the energy gap
to creating a charge $-e/3$ quasiparticle. Therefore, using a gate configuration (see e.g. Fig. \ref{fig1}(c)) it is possible to increase the
amplitude for $t_1$ relative to $t_2$. In fact, since $\log t_1 \propto \epsilon_{e/3}$, the amplitude of $t_1$ can be tuned with exponential
sensitivity in the gate voltage. Similarly, using a gate configuration that follows the loop $\alpha$, (see Fig. \ref{fig2}(b),(c)),
it is possible to increase $t_2$ relative to $t_1$.

The above discussion implies that experiments reminiscent of the fusion rule experiments proposed in Ref. \cite{aasen2015} in the context
of Majorana nanowires can be carried out in this context. For example, suppose the system starts in a limit dominated
by the charging energies, and also $t_2 \gg t_1$. In the limit where the charging energies dominate, the system is in a definite
eigenstate $|n_1\rangle_{\alpha}$. $n_1$ can be determined by measuring the charge in the region surrounded by the loop $\alpha$ (see Fig. \ref{fig1}(b)).
In the opposite limit, where the charging energy is reduced and $t_2$ dominates, the ground state of the system is
of the form $\frac{1}{\sqrt{3}} \sum_{n= 0}^2 e^{2\pi i a n  /3} |n\rangle_{\alpha}$, for some integer $a$. A measurement of the charge in the region will
then give $2ne/3 \text{ mod } 2e$, with probability $1/3$ for each $n$.

\it Analog of non-Abelian braiding without localized zero modes \rm -- Despite the absence of a localized zero mode, it is possible
to implement a robust topological unitary transformation on the topological state space -- an analog of non-Abelian braiding.
The ideas below extend results in \cite{barkeshli2016mcg}, and are closely related to ideas about
braiding non-Abelian defects through tuning interactions \cite{alicea2010b,bonderson2013braiding,clarke2013,lindner2012,barkeshli2013genon,barkeshli2013defect2}.

Let us consider starting with a system with three disconnected boundaries (Fig. \ref{fig2}(b)-(d)). In this case, the system has
Wilson operators $W_1(\alpha_i)$, for $i =1,2$, $W_2(\beta_i)$, for $i=1,2$, and $W_2(\gamma) = W_2(\beta_1)W_2^\dagger(\beta_2)$,
which generalize the operators discussed in the previous section. The states of the system can then be labelled as
$|n, m\rangle_{\beta_1,\beta_2}$, which are eigenstates of $W_2(\beta_i)$, for $i=1,2$, and where $n,m = 0,1,2$.

We consider a Hamiltonian $H(\tau)$ evolving adiabatically in time $\tau$ (measured in units of an overall time scale $T$),
with ground state $|\psi(\tau)\rangle$. We require $H(0) = -t_0 W_2(\beta_2) + H.c.$,
$H(1) = -t_1 W_1(\alpha_2) + H.c.$, $H(2) = -t_2 e^{2\pi i/3} W_1(\alpha_2) W_2(\gamma) + H.c.$, $H(3) = H(0)$.
Furthermore, $[H(\tau), W_2(\beta_1)] \approx 0$ for all $\tau$, $[H(\tau), W_1(\alpha_1)] \approx 0$ for $0 \leq \tau \leq 1$,
$[H(\tau), W_1(\alpha_1)W_1(\alpha_2)] \approx 0$ for $1 \leq \tau \leq 2$, $[H(\tau), W_1(\alpha_1)W_2^\dagger(\gamma)] \approx 0$
for $2 \leq \tau \leq 3$. Here $\approx 0$ means up to exponentially small corrections (see Appendix).

In order to realize the above situation, we use gate configurations in the top and bottom layer in order to enhance certain
quasiparticle tunneling processes of interest. Fig. \ref{fig2}(b)-(d) displays the necessary sequence of active gates for the present discussion.
We begin with a configuration where the top and bottom gates shown in Fig. \ref{fig2}(b) are turned on. The top and bottom gates decrease
the energy gap to creating charge $e/3$ quasiparticles in each layer, thus enhancing the process where charge $2e/3$ quasiparticles
tunnel along the line $\beta_2$ from one edge to another. At time $\tau = 1$, the most dominant quasiparticle tunneling process
should be associated with $W_1(\alpha_2)$, which can be engineered by having the active gates correspond to the configuration in
Fig. \ref{fig2}c. At time $\tau = 2$, the active gates are in the configuration shown in Fig. \ref{fig2}d, and finally at time $\tau = 3$ the Hamiltonian
returns to its original configuration at $\tau = 0$. Thus, we have that the most dominant terms in $H(\tau)$ for $\tau = 0, 1, 2, 3$, are
as required above.

In order for the above terms to be the most dominant terms in the Hamiltonian, it is important that the gate configurations
can sufficiently diminish the energy gap to creating $e/3$ quasiparticles such that even with the increase in geometric length of the paths
(e.g. $\alpha_2 + \gamma$ vs. $\alpha_2)$, the process with the longer path is favorable because of the lower energy gap of the quasiparticle.
Suppose $\epsilon_0$ is the energy gap for creating charge $e/3$ quasiparticles without the gate voltage,
and $\epsilon$ is the energy gap in the presence of the gate voltage. Then, we require, for example,
$\frac{\epsilon}{\epsilon_0} \ll \frac{L_{\alpha_2}}{L_{\alpha_2} + 2 L_\gamma}, \frac{2L_{\gamma}}{L_{\alpha_2} + 2 L_\gamma}$
in order to realize $H(2)$.


The non-Abelian Berry phase of this process gives rise to a unitary transformation $U$ on the topological state space that
we can explicitly compute (see Appendix). The initial state is $|\psi(0)\rangle = \sum_n \psi_{n} |n, b_0\rangle_{\beta_1, \beta_2}$,
for some integer $b_0 = 0,1,2$. One can then verify that $|\psi(3)\rangle = e^{i \vartheta} \sum_n e^{-i \pi (n-k)^2/3 + i \pi (n-k)}|n, b_0\rangle_{\beta_1,\beta_2}$,
where $k$ is an integer that depends on the phases of the tunneling amplitudes $t_i$ for $i=0,1,2$, and $e^{i\vartheta}$ is a non-universal
overall phase. The unitary transformation $U_{mn} = \delta_{mn} e^{i\vartheta} e^{-i \pi (n-k)^2/3 + i \pi (n-k)}$ is identical to
the result for $Z_3$ parafermion zero modes computed in \cite{clarke2013, lindner2012, cheng2012, barkeshli2013genon}.

\it Practical advantages over previous proposals \rm-- The topological degeneracies discussed here
can be induced with only a single type of backscattering term along the edge: the Andreev backscattering term induced by the superconductor.
In contrast, the parafermion zero mode proposals of Ref. \cite{clarke2013,lindner2012,cheng2012} require the ability to gap out the
edge modes in two distinct ways along a single edge, through both magnetically induced normal backscattering and the
superconductivity-induced Andreev backscattering, which may be challenging. Furthermore, the superconducting regions in
our setup should have a diameter $w > \xi_{\text{sc}}$. In contrast, the proposal of \cite{mong2014} requires
superconducting trenches that have a width $w < \xi_{\text{sc}}$, which can be potentially challenging to fabricate.

The proposal of Ref. \cite{barkeshli2014}, where parafermion zero modes are induced through normal interlayer backscattering,
does not admit a way to read out which particular topological state the system is in through an electrostatic charge sensing measurement,
and instead requires quasiparticle  interferometry, which may be more difficult.

\it Generalizations \rm --  The key ingredient in our proposal was the existence of counterpropagating
$\nu  =\pm 1/3$ Laughlin edge modes. The $\nu = 2/3$ FQH state can be viewed as a $\nu = 1$ IQH state,
together with a $\nu = -1/3$ FQH state of holes. Therefore, an electron-electron
bilayer FQH state at total filling fraction $\nu_{\text{total}} = 2/3 + 1/3$ has effectively the same properties
as the $\nu_{\text{total}} = -1/3 + 1/3$ state described here. The main difference is that the $\nu_{\text{total}} = 2/3 + 1/3$ system
will have an additional chiral IQH gapless edge mode along each of the boundaries. Since this edge mode cannot accommodate any
fractional excitations, it has essentially no bearing on the topological degeneracy discussion.

A second generalization of interest is to consider the electron-hole bilayer at $\nu_{\text{total}} = -1/3 + 1/3$, and to consider
normal interlayer electron tunneling $\Psi_1^\dagger \Psi_2 + H.c.$, instead of the superconductivity induced Andreev tunneling.
If the two layers have opposite spin polarization, this can be induced by coupling to a system with either magnetic order or strong spin-orbit coupling, instead of a superconductor.
In this case, much of the discussion of the present paper follows. The main difference is that the topological state cannot be read out
through charge sensing. Naively one might think that the topological state could instead be read out by measuring the electric dipole moment
in a region. However the electric dipole moment is not a conserved quantity. In the limit where the bulk interlayer tunneling is zero
and the superconductor is replaced with a system that spontaneously develops interlayer coherence through interactions, then one could
envision measuring the topological state through fractional electric dipole measurements.

\it Acknowledgments \rm -- I thank Parsa Bonderson, Meng Cheng, Michael Freedman, Chetan Nayak, Javad Shabani,
and Andrea Young for helpful discussions and comments on the manuscript. After this paper was completed, we learned
of work \cite{ganeshan2016} studying the topological degeneracy resulting from electron-hole bilayers with boundaries gapped
by interlayer tunneling.


\begin{thebibliography}{54}
\expandafter\ifx\csname natexlab\endcsname\relax\def\natexlab#1{#1}\fi
\expandafter\ifx\csname bibnamefont\endcsname\relax
  \def\bibnamefont#1{#1}\fi
\expandafter\ifx\csname bibfnamefont\endcsname\relax
  \def\bibfnamefont#1{#1}\fi
\expandafter\ifx\csname citenamefont\endcsname\relax
  \def\citenamefont#1{#1}\fi
\expandafter\ifx\csname url\endcsname\relax
  \def\url#1{\texttt{#1}}\fi
\expandafter\ifx\csname urlprefix\endcsname\relax\def\urlprefix{URL }\fi
\providecommand{\bibinfo}[2]{#2}
\providecommand{\eprint}[2][]{\url{#2}}

\bibitem[{\citenamefont{Nayak et~al.}(2008)\citenamefont{Nayak, Simon, Stern,
  Freedman, and Sarma}}]{nayak2008}
\bibinfo{author}{\bibfnamefont{C.}~\bibnamefont{Nayak}},
  \bibinfo{author}{\bibfnamefont{S.~H.} \bibnamefont{Simon}},
  \bibinfo{author}{\bibfnamefont{A.}~\bibnamefont{Stern}},
  \bibinfo{author}{\bibfnamefont{M.}~\bibnamefont{Freedman}}, \bibnamefont{and}
  \bibinfo{author}{\bibfnamefont{S.~D.} \bibnamefont{Sarma}},
  \bibinfo{journal}{Rev. Mod. Phys.} \textbf{\bibinfo{volume}{80}},
  \bibinfo{pages}{1083} (\bibinfo{year}{2008}).

\bibitem[{\citenamefont{Wen}(1989)}]{wen1989}
\bibinfo{author}{\bibfnamefont{X.~G.} \bibnamefont{Wen}},
  \bibinfo{journal}{Phys. Rev. B} \textbf{\bibinfo{volume}{40}},
  \bibinfo{pages}{7387} (\bibinfo{year}{1989}).

\bibitem[{\citenamefont{Wen and Niu}(1990)}]{wen1990b}
\bibinfo{author}{\bibfnamefont{X.~G.} \bibnamefont{Wen}} \bibnamefont{and}
  \bibinfo{author}{\bibfnamefont{Q.}~\bibnamefont{Niu}},
  \bibinfo{journal}{Phys. Rev. B} \textbf{\bibinfo{volume}{41}},
  \bibinfo{pages}{9377} (\bibinfo{year}{1990}).

\bibitem[{\citenamefont{Moore and Read}(1991)}]{Moore1991}
\bibinfo{author}{\bibfnamefont{G.}~\bibnamefont{Moore}} \bibnamefont{and}
  \bibinfo{author}{\bibfnamefont{N.}~\bibnamefont{Read}},
  \bibinfo{journal}{Nuclear Physics B} \textbf{\bibinfo{volume}{360}},
  \bibinfo{pages}{362 } (\bibinfo{year}{1991}).

\bibitem[{\citenamefont{Wen}(1991)}]{wen1991prl}
\bibinfo{author}{\bibfnamefont{X.~G.} \bibnamefont{Wen}},
  \bibinfo{journal}{Phys. Rev. Lett.} \textbf{\bibinfo{volume}{66}},
  \bibinfo{pages}{802} (\bibinfo{year}{1991}).

\bibitem[{\citenamefont{Nayak and Wilczek}(1996)}]{nayak1996}
\bibinfo{author}{\bibfnamefont{C.}~\bibnamefont{Nayak}} \bibnamefont{and}
  \bibinfo{author}{\bibfnamefont{F.}~\bibnamefont{Wilczek}},
  \bibinfo{journal}{Nuclear Physics B} \textbf{\bibinfo{volume}{479}},
  \bibinfo{pages}{529 } (\bibinfo{year}{1996}), ISSN \bibinfo{issn}{0550-3213}.

\bibitem[{\citenamefont{Bonderson et~al.}(2011)\citenamefont{Bonderson,
  Gurarie, and Nayak}}]{bonderson2011}
\bibinfo{author}{\bibfnamefont{P.}~\bibnamefont{Bonderson}},
  \bibinfo{author}{\bibfnamefont{V.}~\bibnamefont{Gurarie}}, \bibnamefont{and}
  \bibinfo{author}{\bibfnamefont{C.}~\bibnamefont{Nayak}},
  \bibinfo{journal}{Phys. Rev. B} \textbf{\bibinfo{volume}{83}},
  \bibinfo{pages}{075303} (\bibinfo{year}{2011}).

\bibitem[{\citenamefont{Read and Green}(2000)}]{read2000}
\bibinfo{author}{\bibfnamefont{N.}~\bibnamefont{Read}} \bibnamefont{and}
  \bibinfo{author}{\bibfnamefont{D.}~\bibnamefont{Green}},
  \bibinfo{journal}{Phys. Rev. B} \textbf{\bibinfo{volume}{61}},
  \bibinfo{pages}{10267} (\bibinfo{year}{2000}).

\bibitem[{\citenamefont{Kitaev}(2001)}]{kitaev2001}
\bibinfo{author}{\bibfnamefont{A.~Y.} \bibnamefont{Kitaev}},
  \bibinfo{journal}{Physics-Uspekhi} \textbf{\bibinfo{volume}{44}},
  \bibinfo{pages}{131} (\bibinfo{year}{2001}).

\bibitem[{\citenamefont{Fu and Kane}(2008)}]{fu2008}
\bibinfo{author}{\bibfnamefont{L.}~\bibnamefont{Fu}} \bibnamefont{and}
  \bibinfo{author}{\bibfnamefont{C.~L.} \bibnamefont{Kane}},
  \bibinfo{journal}{Phys. Rev. Lett.} \textbf{\bibinfo{volume}{100}},
  \bibinfo{pages}{096407} (\bibinfo{year}{2008}).

\bibitem[{\citenamefont{Sau et~al.}(2010)\citenamefont{Sau, Lutchyn, Tewari,
  and {Das Sarma}}}]{sau2010}
\bibinfo{author}{\bibfnamefont{J.~D.} \bibnamefont{Sau}},
  \bibinfo{author}{\bibfnamefont{R.~M.} \bibnamefont{Lutchyn}},
  \bibinfo{author}{\bibfnamefont{S.}~\bibnamefont{Tewari}}, \bibnamefont{and}
  \bibinfo{author}{\bibfnamefont{S.}~\bibnamefont{{Das Sarma}}},
  \bibinfo{journal}{Phys. Rev. Lett.} \textbf{\bibinfo{volume}{104}},
  \bibinfo{pages}{040502} (\bibinfo{year}{2010}).

\bibitem[{\citenamefont{Alicea}(2010)}]{alicea2010}
\bibinfo{author}{\bibfnamefont{J.}~\bibnamefont{Alicea}},
  \bibinfo{journal}{Phys. Rev. B} \textbf{\bibinfo{volume}{81}},
  \bibinfo{pages}{125318} (\bibinfo{year}{2010}).

\bibitem[{\citenamefont{Lutchyn et~al.}(2010)\citenamefont{Lutchyn, Sau, and
  Das~Sarma}}]{lutchyn2010}
\bibinfo{author}{\bibfnamefont{R.~M.} \bibnamefont{Lutchyn}},
  \bibinfo{author}{\bibfnamefont{J.~D.} \bibnamefont{Sau}}, \bibnamefont{and}
  \bibinfo{author}{\bibfnamefont{S.}~\bibnamefont{Das~Sarma}},
  \bibinfo{journal}{Phys. Rev. Lett.} \textbf{\bibinfo{volume}{105}},
  \bibinfo{pages}{077001} (\bibinfo{year}{2010}).

\bibitem[{\citenamefont{Oreg et~al.}(2010)\citenamefont{Oreg, Refael, and von
  Oppen}}]{oreg2010}
\bibinfo{author}{\bibfnamefont{Y.}~\bibnamefont{Oreg}},
  \bibinfo{author}{\bibfnamefont{G.}~\bibnamefont{Refael}}, \bibnamefont{and}
  \bibinfo{author}{\bibfnamefont{F.}~\bibnamefont{von Oppen}},
  \bibinfo{journal}{Phys. Rev. Lett.} \textbf{\bibinfo{volume}{105}},
  \bibinfo{pages}{177002} (\bibinfo{year}{2010}).

\bibitem[{\citenamefont{Alicea}(2012)}]{alicea2012review}
\bibinfo{author}{\bibfnamefont{J.}~\bibnamefont{Alicea}},
  \bibinfo{journal}{Reports on Progress in Physics}
  \textbf{\bibinfo{volume}{75}}, \bibinfo{pages}{076501}
  (\bibinfo{year}{2012}).

\bibitem[{\citenamefont{Barkeshli and Qi}(2012)}]{barkeshli2012a}
\bibinfo{author}{\bibfnamefont{M.}~\bibnamefont{Barkeshli}} \bibnamefont{and}
  \bibinfo{author}{\bibfnamefont{X.-L.} \bibnamefont{Qi}},
  \bibinfo{journal}{Phys. Rev. X} \textbf{\bibinfo{volume}{2}},
  \bibinfo{pages}{031013} (\bibinfo{year}{2012}), \eprint{arXiv:1112.3311}.

\bibitem[{\citenamefont{You and Wen}(2012)}]{you2012}
\bibinfo{author}{\bibfnamefont{Y.-Z.} \bibnamefont{You}} \bibnamefont{and}
  \bibinfo{author}{\bibfnamefont{X.-G.} \bibnamefont{Wen}},
  \bibinfo{journal}{Phys. Rev. B} \textbf{\bibinfo{volume}{86}},
  \bibinfo{pages}{161107} (\bibinfo{year}{2012}).

\bibitem[{\citenamefont{Bombin}(2010)}]{bombin2010}
\bibinfo{author}{\bibfnamefont{H.}~\bibnamefont{Bombin}},
  \bibinfo{journal}{Phys. Rev. Lett.} \textbf{\bibinfo{volume}{105}},
  \bibinfo{pages}{030403} (\bibinfo{year}{2010}), \eprint{arXiv:1004.1838}.

\bibitem[{\citenamefont{Clarke et~al.}(2013)\citenamefont{Clarke, Alicea, and
  Shtengel}}]{clarke2013}
\bibinfo{author}{\bibfnamefont{D.~J.} \bibnamefont{Clarke}},
  \bibinfo{author}{\bibfnamefont{J.}~\bibnamefont{Alicea}}, \bibnamefont{and}
  \bibinfo{author}{\bibfnamefont{K.}~\bibnamefont{Shtengel}},
  \bibinfo{journal}{Nature Comm.} \textbf{\bibinfo{volume}{4}},
  \bibinfo{pages}{1348} (\bibinfo{year}{2013}), \eprint{arXiv:1204.5479}.

\bibitem[{\citenamefont{Lindner et~al.}(2012)\citenamefont{Lindner, Berg,
  Refael, and Stern}}]{lindner2012}
\bibinfo{author}{\bibfnamefont{N.~H.} \bibnamefont{Lindner}},
  \bibinfo{author}{\bibfnamefont{E.}~\bibnamefont{Berg}},
  \bibinfo{author}{\bibfnamefont{G.}~\bibnamefont{Refael}}, \bibnamefont{and}
  \bibinfo{author}{\bibfnamefont{A.}~\bibnamefont{Stern}},
  \bibinfo{journal}{Phys. Rev. X} \textbf{\bibinfo{volume}{2}},
  \bibinfo{pages}{041002} (\bibinfo{year}{2012}), \eprint{arXiv:1204.5733}.

\bibitem[{\citenamefont{Cheng}(2012)}]{cheng2012}
\bibinfo{author}{\bibfnamefont{M.}~\bibnamefont{Cheng}},
  \bibinfo{journal}{Phys. Rev. B} \textbf{\bibinfo{volume}{86}},
  \bibinfo{pages}{195126} (\bibinfo{year}{2012}), \eprint{arXiv:1204.6084}.

\bibitem[{\citenamefont{Fendley}(2012)}]{fendley2012}
\bibinfo{author}{\bibfnamefont{P.}~\bibnamefont{Fendley}}, \bibinfo{journal}{J.
  Stat. Mech.} p. \bibinfo{pages}{P11020} (\bibinfo{year}{2012}).

\bibitem[{\citenamefont{Kitaev and Kong}(2012)}]{kitaev2012}
\bibinfo{author}{\bibfnamefont{A.}~\bibnamefont{Kitaev}} \bibnamefont{and}
  \bibinfo{author}{\bibfnamefont{L.}~\bibnamefont{Kong}},
  \bibinfo{journal}{Comm. Math. Phys.} \textbf{\bibinfo{volume}{313}},
  \bibinfo{pages}{351} (\bibinfo{year}{2012}).

\bibitem[{\citenamefont{Barkeshli
  et~al.}(2013{\natexlab{a}})\citenamefont{Barkeshli, Jian, and
  Qi}}]{barkeshli2013genon}
\bibinfo{author}{\bibfnamefont{M.}~\bibnamefont{Barkeshli}},
  \bibinfo{author}{\bibfnamefont{C.-M.} \bibnamefont{Jian}}, \bibnamefont{and}
  \bibinfo{author}{\bibfnamefont{X.-L.} \bibnamefont{Qi}},
  \bibinfo{journal}{Phys. Rev. B} \textbf{\bibinfo{volume}{87}},
  \bibinfo{pages}{045130} (\bibinfo{year}{2013}{\natexlab{a}}),
  \eprint{arXiv:1208.4834}.

\bibitem[{\citenamefont{Barkeshli
  et~al.}(2013{\natexlab{b}})\citenamefont{Barkeshli, Jian, and
  Qi}}]{barkeshli2013defect2}
\bibinfo{author}{\bibfnamefont{M.}~\bibnamefont{Barkeshli}},
  \bibinfo{author}{\bibfnamefont{C.-M.} \bibnamefont{Jian}}, \bibnamefont{and}
  \bibinfo{author}{\bibfnamefont{X.-L.} \bibnamefont{Qi}},
  \bibinfo{journal}{Phys. Rev. B} \textbf{\bibinfo{volume}{88}},
  \bibinfo{pages}{235103} (\bibinfo{year}{2013}{\natexlab{b}}).

\bibitem[{\citenamefont{Teo et~al.}(2014)\citenamefont{Teo, Roy, and
  Chen}}]{teo2014}
\bibinfo{author}{\bibfnamefont{J.~C.~Y.} \bibnamefont{Teo}},
  \bibinfo{author}{\bibfnamefont{A.}~\bibnamefont{Roy}}, \bibnamefont{and}
  \bibinfo{author}{\bibfnamefont{X.}~\bibnamefont{Chen}},
  \bibinfo{journal}{Phys. Rev. B} \textbf{\bibinfo{volume}{90}},
  \bibinfo{pages}{155111} (\bibinfo{year}{2014}).

\bibitem[{\citenamefont{Barkeshli
  et~al.}(2014{\natexlab{a}})\citenamefont{Barkeshli, Bonderson, Cheng, and
  Wang}}]{barkeshli2014SDG}
\bibinfo{author}{\bibfnamefont{M.}~\bibnamefont{Barkeshli}},
  \bibinfo{author}{\bibfnamefont{P.}~\bibnamefont{Bonderson}},
  \bibinfo{author}{\bibfnamefont{M.}~\bibnamefont{Cheng}}, \bibnamefont{and}
  \bibinfo{author}{\bibfnamefont{Z.}~\bibnamefont{Wang}}
  (\bibinfo{year}{2014}{\natexlab{a}}), \eprint{arXiv:1410.4540}.

\bibitem[{\citenamefont{Barkeshli and Qi}(2014)}]{barkeshli2014}
\bibinfo{author}{\bibfnamefont{M.}~\bibnamefont{Barkeshli}} \bibnamefont{and}
  \bibinfo{author}{\bibfnamefont{X.-L.} \bibnamefont{Qi}},
  \bibinfo{journal}{Phys. Rev. X} \textbf{\bibinfo{volume}{4}},
  \bibinfo{pages}{041035} (\bibinfo{year}{2014}).

\bibitem[{\citenamefont{Barkeshli
  et~al.}(2014{\natexlab{b}})\citenamefont{Barkeshli, Oreg, and
  Qi}}]{barkeshli2014deg}
\bibinfo{author}{\bibfnamefont{M.}~\bibnamefont{Barkeshli}},
  \bibinfo{author}{\bibfnamefont{Y.}~\bibnamefont{Oreg}}, \bibnamefont{and}
  \bibinfo{author}{\bibfnamefont{X.-L.} \bibnamefont{Qi}}
  (\bibinfo{year}{2014}{\natexlab{b}}), \eprint{arXiv:1401.3750}.

\bibitem[{\citenamefont{Clarke et~al.}(2014)\citenamefont{Clarke, Alicea, and
  Shtengel}}]{clarke2014}
\bibinfo{author}{\bibfnamefont{D.~J.} \bibnamefont{Clarke}},
  \bibinfo{author}{\bibfnamefont{J.}~\bibnamefont{Alicea}}, \bibnamefont{and}
  \bibinfo{author}{\bibfnamefont{K.}~\bibnamefont{Shtengel}},
  \bibinfo{journal}{Nature Physics} \textbf{\bibinfo{volume}{10}},
  \bibinfo{pages}{877} (\bibinfo{year}{2014}).

\bibitem[{\citenamefont{Barkeshli and Nayak}(2015)}]{barkeshli2015fqhsc}
\bibinfo{author}{\bibfnamefont{M.}~\bibnamefont{Barkeshli}} \bibnamefont{and}
  \bibinfo{author}{\bibfnamefont{C.}~\bibnamefont{Nayak}}
  (\bibinfo{year}{2015}), \eprint{arXiv:1507.06305}.

\bibitem[{\citenamefont{Vaezi}(2013{\natexlab{a}})}]{vaezi2013}
\bibinfo{author}{\bibfnamefont{A.}~\bibnamefont{Vaezi}},
  \bibinfo{journal}{Phys. Rev. B} \textbf{\bibinfo{volume}{87}},
  \bibinfo{pages}{035132} (\bibinfo{year}{2013}{\natexlab{a}}).

\bibitem[{\citenamefont{Mong et~al.}(2014)\citenamefont{Mong, Clarke, Alicea,
  Lindner, Fendley, Nayak, Oreg, Stern, Berg, Shtengel et~al.}}]{mong2014}
\bibinfo{author}{\bibfnamefont{R.~S.~K.} \bibnamefont{Mong}},
  \bibinfo{author}{\bibfnamefont{D.~J.} \bibnamefont{Clarke}},
  \bibinfo{author}{\bibfnamefont{J.}~\bibnamefont{Alicea}},
  \bibinfo{author}{\bibfnamefont{N.~H.} \bibnamefont{Lindner}},
  \bibinfo{author}{\bibfnamefont{P.}~\bibnamefont{Fendley}},
  \bibinfo{author}{\bibfnamefont{C.}~\bibnamefont{Nayak}},
  \bibinfo{author}{\bibfnamefont{Y.}~\bibnamefont{Oreg}},
  \bibinfo{author}{\bibfnamefont{A.}~\bibnamefont{Stern}},
  \bibinfo{author}{\bibfnamefont{E.}~\bibnamefont{Berg}},
  \bibinfo{author}{\bibfnamefont{K.}~\bibnamefont{Shtengel}},
  \bibnamefont{et~al.}, \bibinfo{journal}{Phys. Rev. X}
  \textbf{\bibinfo{volume}{4}}, \bibinfo{pages}{011036} (\bibinfo{year}{2014}).

\bibitem[{\citenamefont{Vaezi}(2013{\natexlab{b}})}]{vaezi2013b}
\bibinfo{author}{\bibfnamefont{A.}~\bibnamefont{Vaezi}}
  (\bibinfo{year}{2013}{\natexlab{b}}), \eprint{arXiv:1307.8069}.

\bibitem[{\citenamefont{Vaezi and Barkeshli}(2014)}]{vaezi2014fib}
\bibinfo{author}{\bibfnamefont{A.}~\bibnamefont{Vaezi}} \bibnamefont{and}
  \bibinfo{author}{\bibfnamefont{M.}~\bibnamefont{Barkeshli}},
  \bibinfo{journal}{Phys. Rev. Lett.} \textbf{\bibinfo{volume}{113}},
  \bibinfo{pages}{236804} (\bibinfo{year}{2014}).

\bibitem[{\citenamefont{Barkeshli et~al.}(2015)\citenamefont{Barkeshli, Jiang,
  Thomale, and Qi}}]{barkeshli2015kitaev}
\bibinfo{author}{\bibfnamefont{M.}~\bibnamefont{Barkeshli}},
  \bibinfo{author}{\bibfnamefont{H.-C.} \bibnamefont{Jiang}},
  \bibinfo{author}{\bibfnamefont{R.}~\bibnamefont{Thomale}}, \bibnamefont{and}
  \bibinfo{author}{\bibfnamefont{X.-L.} \bibnamefont{Qi}},
  \bibinfo{journal}{Phys. Rev. Lett.} \textbf{\bibinfo{volume}{114}},
  \bibinfo{pages}{026401} (\bibinfo{year}{2015}).

\bibitem[{\citenamefont{Barkeshli and Freedman}(2016)}]{barkeshli2016mcg}
\bibinfo{author}{\bibfnamefont{M.}~\bibnamefont{Barkeshli}} \bibnamefont{and}
  \bibinfo{author}{\bibfnamefont{M.}~\bibnamefont{Freedman}}
  (\bibinfo{year}{2016}), \eprint{arXiv:1602.01093}.

\bibitem[{\citenamefont{Fowler et~al.}(2012)\citenamefont{Fowler, Mariantoni,
  Martinis, and Cleland}}]{fowler2012}
\bibinfo{author}{\bibfnamefont{A.~G.} \bibnamefont{Fowler}},
  \bibinfo{author}{\bibfnamefont{M.}~\bibnamefont{Mariantoni}},
  \bibinfo{author}{\bibfnamefont{J.~M.} \bibnamefont{Martinis}},
  \bibnamefont{and} \bibinfo{author}{\bibfnamefont{A.~N.}
  \bibnamefont{Cleland}}, \bibinfo{journal}{Phys. Rev. A}
  \textbf{\bibinfo{volume}{86}}, \bibinfo{pages}{032324}
  (\bibinfo{year}{2012}).

\bibitem[{\citenamefont{Wang and Wen}(2015)}]{wang2015}
\bibinfo{author}{\bibfnamefont{J.~C.} \bibnamefont{Wang}} \bibnamefont{and}
  \bibinfo{author}{\bibfnamefont{X.-G.} \bibnamefont{Wen}},
  \bibinfo{journal}{Phys. Rev. B} \textbf{\bibinfo{volume}{91}},
  \bibinfo{pages}{125124} (\bibinfo{year}{2015}), \eprint{arXiv:1212.4863}.

\bibitem[{\citenamefont{Shabani}()}]{shabani}
\bibinfo{author}{\bibfnamefont{J.}~\bibnamefont{Shabani}},
  \bibinfo{note}{private communication}.

\bibitem[{\citenamefont{Abanin et~al.}(2006)\citenamefont{Abanin, Lee, and
  Levitov}}]{abanin2006}
\bibinfo{author}{\bibfnamefont{D.~A.} \bibnamefont{Abanin}},
  \bibinfo{author}{\bibfnamefont{P.~A.} \bibnamefont{Lee}}, \bibnamefont{and}
  \bibinfo{author}{\bibfnamefont{L.~S.} \bibnamefont{Levitov}},
  \bibinfo{journal}{Phys. Rev. Lett.} \textbf{\bibinfo{volume}{96}},
  \bibinfo{pages}{176803} (\bibinfo{year}{2006}).

\bibitem[{\citenamefont{Young}()}]{young}
\bibinfo{author}{\bibfnamefont{A.}~\bibnamefont{Young}}, \bibinfo{note}{private
  communication}.

\bibitem[{\citenamefont{Maher et~al.}(2013)\citenamefont{Maher, Dean, Young,
  Taniguchi, Watanabe, Shepard, Hone, and Kim}}]{maher2013}
\bibinfo{author}{\bibfnamefont{P.}~\bibnamefont{Maher}},
  \bibinfo{author}{\bibfnamefont{C.~R.} \bibnamefont{Dean}},
  \bibinfo{author}{\bibfnamefont{A.~F.} \bibnamefont{Young}},
  \bibinfo{author}{\bibfnamefont{T.}~\bibnamefont{Taniguchi}},
  \bibinfo{author}{\bibfnamefont{K.}~\bibnamefont{Watanabe}},
  \bibinfo{author}{\bibfnamefont{K.~L.} \bibnamefont{Shepard}},
  \bibinfo{author}{\bibfnamefont{J.}~\bibnamefont{Hone}}, \bibnamefont{and}
  \bibinfo{author}{\bibfnamefont{P.}~\bibnamefont{Kim}},
  \bibinfo{journal}{Nature Physics} \textbf{\bibinfo{volume}{9}},
  \bibinfo{pages}{154} (\bibinfo{year}{2013}).

\bibitem[{\citenamefont{Sanchez-Yamagishi
  et~al.}(2016)\citenamefont{Sanchez-Yamagishi, Luo, Young, Hunt, Watanabe,
  Taniguchi, Ashoori, and Jarillo-Herrero}}]{sanchez2016}
\bibinfo{author}{\bibfnamefont{J.~D.} \bibnamefont{Sanchez-Yamagishi}},
  \bibinfo{author}{\bibfnamefont{J.~Y.} \bibnamefont{Luo}},
  \bibinfo{author}{\bibfnamefont{A.~F.} \bibnamefont{Young}},
  \bibinfo{author}{\bibfnamefont{B.}~\bibnamefont{Hunt}},
  \bibinfo{author}{\bibfnamefont{K.}~\bibnamefont{Watanabe}},
  \bibinfo{author}{\bibfnamefont{T.}~\bibnamefont{Taniguchi}},
  \bibinfo{author}{\bibfnamefont{R.~C.} \bibnamefont{Ashoori}},
  \bibnamefont{and}
  \bibinfo{author}{\bibfnamefont{P.}~\bibnamefont{Jarillo-Herrero}}
  (\bibinfo{year}{2016}), \eprint{arXiv:1602.06815}.

\bibitem[{\citenamefont{Wan et~al.}(2015)\citenamefont{Wan, Kazakov, Manfra,
  Pfeiffer, West, and Rokhinson}}]{rokhinson2015}
\bibinfo{author}{\bibfnamefont{Z.}~\bibnamefont{Wan}},
  \bibinfo{author}{\bibfnamefont{A.}~\bibnamefont{Kazakov}},
  \bibinfo{author}{\bibfnamefont{M.~J.} \bibnamefont{Manfra}},
  \bibinfo{author}{\bibfnamefont{L.~N.} \bibnamefont{Pfeiffer}},
  \bibinfo{author}{\bibfnamefont{K.~W.} \bibnamefont{West}}, \bibnamefont{and}
  \bibinfo{author}{\bibfnamefont{L.~P.} \bibnamefont{Rokhinson}}
  (\bibinfo{year}{2015}), \eprint{arXiv:1503.09138}.

\bibitem[{\citenamefont{Shabani et~al.}(2015)\citenamefont{Shabani, Kjaergaard,
  Suominen, Kim, Nichele, Pakrouski, Stankevic, Lutchyn, Krogstrup,
  Feidenhans'l et~al.}}]{shabani2015}
\bibinfo{author}{\bibfnamefont{J.}~\bibnamefont{Shabani}},
  \bibinfo{author}{\bibfnamefont{M.}~\bibnamefont{Kjaergaard}},
  \bibinfo{author}{\bibfnamefont{H.~J.} \bibnamefont{Suominen}},
  \bibinfo{author}{\bibfnamefont{Y.}~\bibnamefont{Kim}},
  \bibinfo{author}{\bibfnamefont{F.}~\bibnamefont{Nichele}},
  \bibinfo{author}{\bibfnamefont{K.}~\bibnamefont{Pakrouski}},
  \bibinfo{author}{\bibfnamefont{T.}~\bibnamefont{Stankevic}},
  \bibinfo{author}{\bibfnamefont{R.~M.} \bibnamefont{Lutchyn}},
  \bibinfo{author}{\bibfnamefont{P.}~\bibnamefont{Krogstrup}},
  \bibinfo{author}{\bibfnamefont{R.}~\bibnamefont{Feidenhans'l}},
  \bibnamefont{et~al.} (\bibinfo{year}{2015}), \eprint{arXiv:1511.01127}.

\bibitem[{\citenamefont{Amet et~al.}(2015)\citenamefont{Amet, Ke, Borzenets,
  Wang, Watanabe, Taniguchi, Deacon, Yamamoto, Bomze, Tarucha
  et~al.}}]{finkelstein2015}
\bibinfo{author}{\bibfnamefont{F.}~\bibnamefont{Amet}},
  \bibinfo{author}{\bibfnamefont{C.~T.} \bibnamefont{Ke}},
  \bibinfo{author}{\bibfnamefont{I.~V.} \bibnamefont{Borzenets}},
  \bibinfo{author}{\bibfnamefont{Y.-M.} \bibnamefont{Wang}},
  \bibinfo{author}{\bibfnamefont{K.}~\bibnamefont{Watanabe}},
  \bibinfo{author}{\bibfnamefont{T.}~\bibnamefont{Taniguchi}},
  \bibinfo{author}{\bibfnamefont{R.~S.} \bibnamefont{Deacon}},
  \bibinfo{author}{\bibfnamefont{M.}~\bibnamefont{Yamamoto}},
  \bibinfo{author}{\bibfnamefont{Y.}~\bibnamefont{Bomze}},
  \bibinfo{author}{\bibfnamefont{S.}~\bibnamefont{Tarucha}},
  \bibnamefont{et~al.} (\bibinfo{year}{2015}), \eprint{arXiv:1512.09083}.

\bibitem[{\citenamefont{Wen}(1995)}]{wen1995}
\bibinfo{author}{\bibfnamefont{X.-G.} \bibnamefont{Wen}},
  \bibinfo{journal}{Adv. Phys.} \textbf{\bibinfo{volume}{44}},
  \bibinfo{pages}{405} (\bibinfo{year}{1995}).

\bibitem[{\citenamefont{Senthil and Fisher}(2001)}]{senthil2001}
\bibinfo{author}{\bibfnamefont{T.}~\bibnamefont{Senthil}} \bibnamefont{and}
  \bibinfo{author}{\bibfnamefont{M.~P.~A.} \bibnamefont{Fisher}},
  \bibinfo{journal}{Phys. Rev. B} \textbf{\bibinfo{volume}{64}},
  \bibinfo{pages}{214511} (\bibinfo{year}{2001}).

\bibitem[{\citenamefont{Venkatachalam et~al.}(2011)\citenamefont{Venkatachalam,
  Yacoby, Pfeiffer, and West}}]{venkatachalam2011}
\bibinfo{author}{\bibfnamefont{V.}~\bibnamefont{Venkatachalam}},
  \bibinfo{author}{\bibfnamefont{A.}~\bibnamefont{Yacoby}},
  \bibinfo{author}{\bibfnamefont{L.}~\bibnamefont{Pfeiffer}}, \bibnamefont{and}
  \bibinfo{author}{\bibfnamefont{K.}~\bibnamefont{West}},
  \bibinfo{journal}{Nature} \textbf{\bibinfo{volume}{469}},
  \bibinfo{pages}{185} (\bibinfo{year}{2011}).

\bibitem[{\citenamefont{Aasen et~al.}(2015)\citenamefont{Aasen, Hell, Mishmash,
  Higginbotham, Danon, Leijnse, Jespersen, Folk, Marcus, and
  Karsten~Flensberg}}]{aasen2015}
\bibinfo{author}{\bibfnamefont{D.}~\bibnamefont{Aasen}},
  \bibinfo{author}{\bibfnamefont{M.}~\bibnamefont{Hell}},
  \bibinfo{author}{\bibfnamefont{R.~V.} \bibnamefont{Mishmash}},
  \bibinfo{author}{\bibfnamefont{A.}~\bibnamefont{Higginbotham}},
  \bibinfo{author}{\bibfnamefont{J.}~\bibnamefont{Danon}},
  \bibinfo{author}{\bibfnamefont{M.}~\bibnamefont{Leijnse}},
  \bibinfo{author}{\bibfnamefont{T.~S.} \bibnamefont{Jespersen}},
  \bibinfo{author}{\bibfnamefont{J.~A.} \bibnamefont{Folk}},
  \bibinfo{author}{\bibfnamefont{C.~M.} \bibnamefont{Marcus}},
  \bibnamefont{and} \bibinfo{author}{\bibfnamefont{J.~A.}
  \bibnamefont{Karsten~Flensberg}} (\bibinfo{year}{2015}),
  \eprint{arXiv:1511.05153}.

\bibitem[{\citenamefont{Alicea et~al.}(2010)\citenamefont{Alicea, Oreg, Refael,
  von Oppen, and Fisher}}]{alicea2010b}
\bibinfo{author}{\bibfnamefont{J.}~\bibnamefont{Alicea}},
  \bibinfo{author}{\bibfnamefont{Y.}~\bibnamefont{Oreg}},
  \bibinfo{author}{\bibfnamefont{G.}~\bibnamefont{Refael}},
  \bibinfo{author}{\bibfnamefont{F.}~\bibnamefont{von Oppen}},
  \bibnamefont{and} \bibinfo{author}{\bibfnamefont{M.~P.~A.}
  \bibnamefont{Fisher}}, \bibinfo{journal}{Nature Physics}
  \textbf{\bibinfo{volume}{7}}, \bibinfo{pages}{412} (\bibinfo{year}{2010}).

\bibitem[{\citenamefont{Bonderson}(2013)}]{bonderson2013braiding}
\bibinfo{author}{\bibfnamefont{P.}~\bibnamefont{Bonderson}},
  \bibinfo{journal}{Phys. Rev. B} \textbf{\bibinfo{volume}{87}},
  \bibinfo{pages}{035113} (\bibinfo{year}{2013}).

\bibitem[{\citenamefont{Ganeshan et~al.}()\citenamefont{Ganeshan, Gorshkov,
  Gurarie, and Galitski}}]{ganeshan2016}
\bibinfo{author}{\bibfnamefont{S.}~\bibnamefont{Ganeshan}},
  \bibinfo{author}{\bibfnamefont{A.}~\bibnamefont{Gorshkov}},
  \bibinfo{author}{\bibfnamefont{V.}~\bibnamefont{Gurarie}}, \bibnamefont{and}
  \bibinfo{author}{\bibfnamefont{V.}~\bibnamefont{Galitski}},
  \bibinfo{note}{unpublished}.

\end{thebibliography}

\newpage
\appendix

\begin{widetext}
\section{Appendix: Non-Abelian Berry phase calculation and discussion}

Let us consider starting with a system with three disconnected boundaries, as shown in Fig. \ref{fig2} (b)-(d) of the main text.
The Hamiltonian evolution $H(\tau)$ splits the 9-dimensional subspace into a three-dimensional ground state subspace, with a finite
energy gap $\delta E$ to the rest of the states. $\delta E$ must remain finite for the entire process, and sets the time scale for the
adiabaticity of $H(\tau)$.

$H(\tau)$ further has the following properties:
\begin{align}
\label{targets}
H(0) &= -t_0 W_2(\beta_2) + H.c.
\nonumber \\
H(1) &= -t_1 W_1(\alpha_2) + H.c.
\nonumber \\
H(2) &= -t_2 e^{2\pi i /3} W_1(\alpha_2)W_2(\gamma) + H.c.
\nonumber \\
H(3) &= -t_0 W_2(\beta_2) + H.c.
\end{align}
Moreover, we require that
\begin{align}
\label{comrel}
[W_2(\beta_1), H(\tau)] &=0, \;\; \forall \tau,
\nonumber \\
[W_1(\alpha_1), H(\tau)] &=0 \;\; \text{ for } 0 \leq \tau \leq 1,
\nonumber \\
[W_1(\alpha_1)W_1(\alpha_2), H(\tau)] &=0 \;\; \text{ for } 1 \leq \tau \leq 2,
\nonumber \\
[W_1(\alpha_1)W^\dagger(\gamma), H(\tau)] &=0 \;\; \text{ for } 2 \leq \tau \leq 3 .
\end{align}
The non-Abelian Berry phase that we compute below is exact and independent of any other details of the adiabatic evolution of $H(\tau)$,
as long as the above conditions are met. The topologically robustness of the non-Abelian Berry phase is derived from the ability
to satisfy the above equations, up to exponentially small corrections in physically controllable parameters,
as we discuss below.

\subsection{Basis choices}

Since the system has three disconnected boundaries gapped by superconductivity, the topological space of states is
$9$-dimensional. The effective Hamiltonian $H(\tau)$ splits this $9$-dimensional ground state subspace into a
three-dimensional ground state subspace with a finite energy gap to the rest of the states.
It will be useful to consider three different basis choices, which we will label as
$|a,b\rangle_{L_I}$, for $I = 1,2,3$. Here $L_I$ denote a set of paths:
\begin{align}
L_1 &= \{(\beta_1,\alpha_1), (\beta_2, \alpha_2)\}
\nonumber \\
L_2 &= \{(\beta_1, \alpha_1 + \alpha_2), (\alpha_2 + \gamma, \alpha_2)\}
\nonumber \\
L_3 &= \{ (\beta_1, \alpha_1 - \gamma), (\beta_2, \alpha_2 + \gamma)\} .
\end{align}

If we define a set of operators $A_i, B_i$ for $i = 1,2$, such that
\begin{align}
\label{magAlg}
A_i B_j &= e^{\delta_{ij} 2\pi i/3}B_j A_i
\nonumber \\
[A_i, A_j] &= [B_i, B_j] = 0,
\end{align}
then we can set
\begin{align}
(B_{i}^{(1)}, A_{i}^{(1)}) = (W_2(\beta_i), W_1(\alpha_i)),
\end{align}
and the algebra of Eq. (\ref{magAlg}) will be satisfied. Alternatively, we can set
\begin{align}
\{(B_{1}^{(2)}, A_1^{(2)}), (B_2^{(2)}, A_2^{(2)})\} = \{ (W_2(\beta_1), W_1(\alpha_1)W_1(\alpha_2)), ( e^{-2\pi i /3} W_1^\dagger (\alpha_2) W_2^\dagger(\gamma), W_1(\alpha_2)) \},
\end{align}
and the algebra of Eq. (\ref{magAlg}) will be satisfied. Finally, we can set
\begin{align}
\{(B_1^{(3)}, A_1^{(3)}, (B_2^{(3)}, A_2^{(3)}\} = \{ (W_2(\beta_1), W_1(\alpha_1)W_2^\dagger(\gamma)), (W_2(\beta_2), W_1(\alpha_2)W_2(\gamma) ) \}.
\end{align}

The three different bases that we are interested in have the following property:
\begin{align}
A^{(I)}_1 | a_1, a_2\rangle_{L_I} = |(a_1 + 1) \% 3, a_2\rangle_{L_I}
\nonumber \\
A^{(I)}_2 | a_1, a_2\rangle_{L_I} = |a_1, (a_2 + 1 )\% 3\rangle_{L_I}
\nonumber \\
B^{(I)}_i | a_1, a_2\rangle_{L_I} = e^{2\pi i a_i/3} |a_1, a_2 \rangle_{L_I}
\end{align}

Now we observe that the different bases can be related to each other as follows:
\begin{align}
|a, b \rangle_{L_2} &= \frac{1}{\sqrt{3}} \sum_{n'=0}^2 e^{(n' - a - b)^2 2\pi i /3} |a, n'\rangle_{L_1}
\nonumber \\
|n, m\rangle_{L_1} &= \frac{1}{\sqrt{3}} \sum_{b=0}^2 e^{- (b+n-m)^2 2\pi i /3} |n,b \rangle_{L_2} .
\end{align}

Furthermore,
\begin{align}
|a, b\rangle_{L_3} &= e^{ ((a - b)^2 + a) 2\pi i /3} |a, b\rangle_{L_1},
\nonumber \\
|a, b\rangle_{L_1} &= e ^{- ((a-b)^2 + a) 2\pi i /3} |a, b\rangle_{L_3}
\end{align}

\section{Non-Abelian Berry phase result}

Now we are in a position to determine the result of the non-Abelian Berry phase resulting from the Hamiltonian evolution $H(\tau)$.
The ground state wave function at $\tau = 0$ is
\begin{align}
|\psi(0) \rangle = \sum_n \psi_n |n, b_0 \rangle_{L_1} .
\end{align}
The integer $b_0 = 0,1,2$ is determined by the tunneling amplitude $t_0$. The ground state energy of $H(0)$ is
$E(0) = -2 |t_0| \cos(2\pi b_0 /3 + \theta_0)$, where $t_0 = |t_0| e^{i\theta}$. The integer $b_0$ is thus chosen to minimize
$E(0)$. We see that when $t_0$ is real and positive, $b_0 = 0$.

The state $|\psi(1)\rangle$ is a ground state of $H(1)$, and is therefore of the form
\begin{align}
|\psi(1)\rangle &= e^{i \phi} \frac{1}{\sqrt{3}} \sum_{n,a} \psi_n e^{2\pi i b_1 a/3} |n, a\rangle_{L_1}.
\end{align}
Again, the integer $b_1 = 0,1,2$ depends on the tunneling amplitudes $t_1$ such that the ground
state energy $E(1) = -2 |t_1| \cos(2\pi b_1/3 + \theta_1)$ is minimized.
Crucially, the Berry phase $e^{i\phi}$ is independent of $n$. Whereas this would not be true in the
most general possible evolution from $H(0)$ to $H(1)$, it is true in our case because of the fact that
$[W_2(\beta_1), H(\tau)] = [W_1(\alpha_1), H(\tau)] = 0$ for $0 \leq \tau \leq 1$.
These commutation relations imply that the Berry phase $e^{i\phi}$ must be independent of the integer $n$, because
there is no term in $H(\tau)$ from $\tau = 0$ to $1$, which can distinguish states with different $n$. Only operators
generated by $W_2(\beta_1)$, $W_1(\alpha_1)$ can distinguish states with different $n$, and they do not appear in
$H(\tau)$ for $\tau = 0$ to $1$. We will use this argument repeatedly below for the subsequent steps as well.

To proceed, we rewrite $|\psi(1)\rangle$ in the $L_2$ basis:
\begin{align}
|\psi(1)\rangle &= e^{i \phi} \frac{1}{3} \sum_{n,a} \psi_n e^{2\pi i b_1 a/3}\sum_c e^{- (c+n-a)^2 2\pi i /3} |n,c \rangle_{L_2}
\nonumber \\
&= -i e^{i\phi} \frac{1}{\sqrt{3}}e^{ 2\pi i b_1^2 /3} \sum_{n} e^{2\pi i b_1 n/3} \psi_n \sum_c e^{ 2\pi i b_1 c/3} |n, c\rangle_{L_2} .
\end{align}

$|\psi(2)\rangle$ is a ground state of $H(2)$, and is of the form
\begin{align}
|\psi(2) \rangle &= e^{i \phi'} \sum_n e^{2\pi i b_1 n/3} \psi_n |n, b_2 \rangle_{L_2}
\end{align}
The integer $b_2 =0,1,2$ depends on the ground state of $E(2)$, similar to the case with $b_0$, $b_1$. For $t_2$ real and positive, $b_2 = 0$.
Again, similar to the previous case, the crucial point here is that $e^{i\phi'}$ is independent of $n$, because
$[W_1(\beta_1), H(\tau)] = [W_1(\alpha_1)W_1(\alpha_2), H(\tau)] = 0$ for $1 \leq \tau \leq 2$, and operators
generated by $W_1(\beta_1)$, $W_1(\alpha_1)W_1(\alpha_2)$ are the only operators that, in the $L_2$ basis, can distinguish states with
different $n$.

To proceed to the next step, we rewrite $|\psi(2)\rangle$ in the $L_3$ basis:
\begin{align}
|\psi(2) \rangle &= e^{i \phi'} \sum_n e^{2\pi i b_1 n/3}\psi_n |n, b_2 \rangle_{L_2}
\nonumber \\
&= e^{i\phi'} \frac{1}{\sqrt{3}} \sum_n e^{2\pi i b_1 n/3}\psi_n \sum_{n'} e^{2\pi i (n' - n -b_2)^2/3} |n, n'\rangle_{L_1}
\nonumber \\
&= e^{i\phi'} \frac{1}{\sqrt{3}} \sum_n e^{2\pi i b_1 n/3}\psi_n \sum_{n'} e^{2\pi i (n' - n -b_2)^2/3} e^{-2\pi i ((n -n' )^2 + n)/3} |n, n'\rangle_{L_3}
\nonumber \\
&= e^{i\phi'} \frac{1}{\sqrt{3}} e^{2\pi i b_2^2/3} \sum_n e^{2\pi i (b_1-b_2 -1) n/3}\psi_n \sum_{n'} e^{2\pi i n' b_2/3 } |n, n'\rangle_{L_3}
\end{align}

Finally, we see that $|\psi(3)\rangle$ is a ground state of $H(3)$, and is then of the form
\begin{align}
|\psi(3)\rangle &=
e^{i\phi''} e^{2\pi i b_2^2/3} \sum_n e^{2\pi i (b_1-b_2 -1) n/3}\psi_n  e^{2\pi i b_0 b_2/3 } |n, b_0\rangle_{L_3}
\nonumber \\
&= e^{i\phi''} e^{2\pi i b_2^2/3} \sum_n e^{2\pi i (b_1-b_2 -1) n/3}\psi_n  e^{2\pi i b_0 b_2/3 } e^{2\pi i/3 ((n-b_0)^2 + n)}|n, b_0\rangle_{L_1}
\nonumber \\
&= e^{i\phi''} e^{2\pi i/3 b_1 (b_0 - b_2 - b_1)} \sum_n \psi_n e^{2\pi i/3 (n-b_0+b_2-b_1)^2}|n, b_0\rangle_{L_1}
\end{align}
Again, the overall phase $e^{i\phi''}$ is the Berry phase obtained in adiabatically passing from $H(2)$ to $H(3)$. $\phi''$ must be independent
of $n$ because $H(\tau)$ for $2 \leq \tau \leq 3$ commutes with all operators that can distinguish states associated with different $n$ (in the $L_3$ basis).

We therefore see that the adiabatic process gave rise to the transformation:
\begin{align}
|\psi(0)\rangle = \sum_n \psi_n |n, b_0\rangle_{L_1} \rightarrow |\psi(3)\rangle = e^{i \vartheta} \sum_n \psi_n e^{2\pi i/3 (n-k)^2}|n, b_0\rangle_{L_1},
\end{align}
where $k = b_0 + b_1 - b_2$, and $\vartheta$ is a non-topological, path-dependent contribution to the overall phase.
In other words, $H(\tau)$ implemented a unitary transformation $U$ as a non-Abelian Berry phase:
\begin{align}
U_{nm} = \delta_{nm} e^{i\vartheta} e^{2\pi i /3 (n-k)^2} = \delta_{nm} e^{i\vartheta} e^{-i \pi (n-k)^2/3 + i \pi (n-k)}
\end{align}

\section{Topological Protection}

The topological protection of the above adiabatic process derives from the ability to satisfy the conditions described in
Eq. (\ref{targets})-(\ref{comrel}), up to exponentially small corrections. Usually,
in the context of braiding of non-Abelian anyons, the corrections to braiding can be made exponentially small by spatially
separating the anyons arbitrarily far apart from each other compared to the correlation length of the system.

In the presence case, the way to achieve the conditions required is to enhance the tunneling amplitudes of quasiparticles
along the paths of interest. Let $\omega$ and $\omega'$ denote two paths, and $L_\omega$, $L_{\omega'}$ their lengths. The
tunneling amplitudes for charge $e/3$ quasiparticles along these paths are $t_\omega \propto e^{- \epsilon_{e/3} L_{\omega}}$,
$t_{\omega'} \propto e^{- \epsilon_{e/3} L_{\omega'}}$. We see that $t_\omega$ can be made exponentially larger than $t_{\omega'}$
as long as the energy gap to created $e/3$ quasiparticles along $\omega$ can be made to be much less than the corresponding gap
along the path $\omega'$. The ratio $t_{\omega'}/t_{\omega} \rightarrow 0$ in the limit that the
$\epsilon_{e/3}^{(\omega)}/\epsilon_{e/3}^{(\omega')} \rightarrow 0$, where $\epsilon^{\omega}_{e/3}$ is the energy gap to creating
$e/3$ quasiparticles along the path $\omega$.

Thus, the topological protection of the non-Abelian Berry phase results from the possibility of controlling the energy
gaps of quasiparticles along paths of interest to be parametrically smaller than the energy gaps along other paths, thus exponentially
enhancing the amplitude for desired tunneling paths over others. In this way, the topological robustness of the
non-Abelian Berry phase obtained here is somewhat different from that of braiding non-Abelian anyons, as the latter can
approach the topological limit with exponentially small corrections by increasing the distance between the anyons.

\end{widetext}

\end{document}